 \documentclass[reprint]{JASAnew}

\newcommand{\mi}{\mathrm{i}}
\newcommand{\pfrac}[2]{\frac{\partial #1 }{\partial #2 }}
\newcommand{\lr}[1]{\left( #1 \right)}
\newcommand{\ty}{  Y }
\newcommand{\ceil}[1]{\lceil #1 \rceil}
\newcommand{\intd}{\mathrm{d}}
\newcommand{\ex}{\textnormal{e}}
\newcommand{\F}{\mathbb{F}}
\newcommand{\iF}{\mathbb{F}^{-1}}

\usepackage[normalem]{ulem}

\newcommand{\Rev}[2]{#2}

\begin{document}

\title[ JASA/ Time-fractional static memory power-law absorption]{Modelling power-law ultrasound absorption using a time-fractional, static memory, Fourier pseudo-spectral method}
\author{Matthew J. King}
\email{matthew.king@ucl.ac.uk}
\affiliation{Department of Medical Physics and Biomedical Engineering,  University College London, London, WC1E 6BT,
United Kingdom}
\author{Timon S. Gutleb}
\affiliation{School of Computer Science, University of Leeds, Leeds, LS2 9JT, United Kingdom}

\author{B. E. Treeby}
\author{B. T. Cox}	
\affiliation{Department of Medical Physics and Biomedical Engineering,  University College London, London, WC1E 6BT,
United Kingdom}

\preprint{Author, JASA}		

\date{14 March 2025} 

\begin{abstract}
We describe and implement a numerical method \Rev{for evaluating a numerical method}{} for modelling the frequency-dependent power-law absorption of ultrasound in tissue\Rev{ using}{, as governed by} the first order linear wave equations with a loss taking the form of a fractional time derivative. The (Caputo) fractional time derivative requires the full problem history which is contained within an iterative procedure.  \Rev{with the}{The} resulting numerical method requires a \Rev{static memory at across all time steps}{fixed (static) memory cost irrespective of the number of time steps.}\Rev{without loss of accuracy.}{}  The spatial domain is treated by the Fourier spectral method. Numerically comparisons are made against a model for the same power-law absorption with loss described by the fractional-Laplacian operator. One advantage of the fractional time derivative over the fractional-Laplacian operator is the local treatment of the power-law, allowing for a spatially varying frequency power-law.  This article may be downloaded for personal use only. Any other use requires prior permission of the author and the Acoustical Society of America. The following article appeared in J. Acoust. Soc. Am. 1 March 2025; 157 (3): 1761–1771 and may be found at https://pubs.aip.org/asa/jasa/article/157/3/1761/3339551
\end{abstract}

\maketitle

\section{Introduction\label{sec:1}}

The rate of absorption of ultrasound waves in the body depends on frequency and, for the range of frequencies of interest for biomedical ultrasound, typically follows a power-law \citep{wells1975absorption,narayana1983frequency,goss1979ultrasonic,treeby2010modeling}. The absorption coefficient is therefore commonly written in the form $\alpha = \alpha_0\omega^y$, where $\omega$ is the temporal frequency and $\alpha_0$ and $y$ are tissue-dependent parameters \cite{szabo1994time,chen2004fractional}. Lossy wave equations that account for frequency-dependent absorption were first derived for integer values of the exponent $y$ \cite{blackstock1967transient,stokes1845theories}. However, for most biological tissues $y$ is a non-integer lying in the range $0<y<2$ \cite{li20213}. 
Wave equations exhibiting this behaviour have been developed using fractional time derivative operators \cite{caputo1967linear,szabo1994time,liebler2004full,wismer2006finite}. 
A consequence of the 
use of a time-fractional derivative on numerical approaches to solving these equations is the need to store some history of the wavefield, and hence the need for additional computational memory. The amount of additional memory required grows significantly as $y$ approaches 1.
To avoid this memory issue a fractional-Laplacian has been used in place of a fractional time derivative, \cite{chen2003physical,treeby2010modeling}. This can be straightforwardly computed within the context of pseudo-spectral models using the Fast Fourier Transform (FFT) without the need to store the history of the wavefield. This is the approach taken in the software package k-Wave (see http://www.k-wave.org)\cite{treeby2010k}. However, there are drawbacks. First, the power-law that is modelled is not strictly a power-law of the frequency, $\omega^{y}$ \Rev{but of the wave-number, $k^{y}$, and second,}{as would be observed in experiments, but of the wave-number $k^{y}$ which for $\omega$ to follow the power law relies on close adherence to the loss-less dispersion relation $\omega=c_0 k$. Secondly},
the power-law exponent $y$ - because it is implemented in k-space - cannot be spatially-varying. \Rev{It is therefore necessary to approximate the heterogeneous absorption that will be present in practice in tissue using a single power-law, eg. by applying a homogenisation technique \cite{li20213}.}{}
To allow heterogeneous power-law absorption to be modelled accurately, this paper \Rev{therefore}{} focuses on the equations that use fractional time derivatives and looks at alternative methods of tackling the memory problem. 

Several alternative approaches for resolving the increased memory requirement of time-fractional derivative methods have been suggested, including the fixed memory principle \cite{podlubny1998fractional} and the logarimthmic memory principle \cite{ford2001numerical}, as well as methods  \cite{yuan2002numerical,diethelm2008investigation,birk2010improved} that re-frame the integral into a double integral form, accounting for the full history with an iterative procedure and quadrature rule. 
The specific procedure used here is of this latter type, adapted from the work of \cite{birk2010improved}.
This method is also discussed and a range of numerical examples are given in \cite{gutleb2023static}, which highlights specific implementation approaches that we adapt to make use of the k-space pseudo-spectral method and minimise the additional memory costs of the approach. \citet{gutleb2023static} additionally discuss the potential for the time-fractional derivative given being used on lossy wave equation given by Caputo/Wismer, citing this as their primary motivation. However, they actually make use of the form given by Szabo \cite{szabo1994time}. This is performed on a  two-dimensional (2D) disk in line with the majority of \cite{gutleb2023static} focusing on a spectral approach in order to resolve the problems in a wider range of spatial domains.

This work is divided into two main sections. Section \ref{sec:2} gives the theoretical background, 
the governing equations and each of their parameters, along with a breakdown of the static-memory method for the fractional time derivative. It concludes with a discussion of the numerical scheme used for the evaluation of the fractional time derivatives.
In Section \ref{sec:3} a number of numerical results are presented illustrating the convergence of the static memory method in each of its parameters with comparisons to the fractional-Laplacian approach (as implemented in k-Wave). This is followed by the illustration of the ability of the time-fractional method to correctly predict the absorption through a one-dimensional (1D) example. Section \ref{sec:3} concludes by considering examples with heterogeneous $y$ in both 1D and three dimensional (3D), making comparisons to a homogenised k-Wave formulation.\\

\section{\label{sec:2} Theory}
\subsection{Governing equations}
One way to model power-law absorption of acoustic pressure waves is through a wave equation with a loss term described, as in the Caputo/Wismer equations, by the fractional time derivative of the Laplacian of the pressure $p$:
\begin{equation}\frac{1}{c_0^2}\pfrac{^2p}{t^2}=\nabla^2 p +\tau \pfrac{^{y-1}}{t^{y-1}}\nabla^2 p,\label{SecondOrder}\end{equation}
where $c_0$ is the sound speed and $\tau$ is related to the absorption coefficient.
However, instead of solving this equation directly, we will instead make use of the equivalent system of first-order equations:
\begin{subequations}\label{firstorder}
\begin{align} \rho_0 \pfrac{\boldsymbol{u}}{t} &= -\boldsymbol{\nabla} p, \label{CoMo}\\
\pfrac{\rho}{t} &= -\rho_0 \boldsymbol{\nabla} \cdot \boldsymbol{u}, \label{CoMa}\\
p&=c_0^2 \lr{ \rho + \tau \pfrac{^{y-1}\rho}{t^{y-1}} }. \label{State} \end{align}
\end{subequations} 
\Rev{}{The introduction of the time-fractional derivative in the acoustic equation of state \eqref{State} may be viewed as accounting for the material properties, and can be interpreted physically in terms of springs and dampers \cite{mainardi2022fractional}.}
By taking the second time derivative of \eqref{State} it is easy to show that these equations give rise to \eqref{SecondOrder}.
An expression for $\tau$ in terms of the absorption coefficient can be found by considering the propagation of the plane wave.
\begin{equation}\label{plane}
    p=p_0\exp^{\mi(kx-\omega t)}.
\end{equation}
    \Rev{It is noted that}{We note that} the wave will only attenuate for a real frequency $\omega$ when the wavenumber $k=k_r+\mi k_i$ has a positive imaginary part, with the decay rate\Rev{dictated by the value of}{} $k_i$. Since we are modelling power-law absorption following $\alpha=\alpha_0 \omega^y$, we set $k_i=\alpha$. Expanding \Rev{ \eqref{firstorder}, or equivalently}{} \eqref{SecondOrder} with \eqref{plane}, gives;
\begin{align}
    k_i^2 =& \frac{1}{2}\frac{\omega^2}{c_0^2}\frac{-\bigg(1+\tau\omega^{y-1}\cos\big(\frac{-\pi(y-1)}{2}\big)\bigg)}{1+2\tau\omega^{y-1}\cos\big(\frac{-\pi(y-1)}{2}\big) + \big(\tau\omega^{y-1}\big)^2}\nonumber \\
    & \phantom{\frac{1}{2}}\pm\frac{ \sqrt{ 1+2\tau\omega^{y-1}\cos\big(\frac{-\pi(y-1)}{2}\big) + \big(\tau\omega^{y-1}\big)^2  }}{1+2\tau\omega^{y-1}\cos\big(\frac{-\pi(y-1)}{2}\big) + \big(\tau\omega^{y-1}\big)^2}.
\end{align}
While the above expression is exact, if we additionally assume that $\tau\omega^{y-1}\ll 1$ the resulting asymptotic expression for $k_i$ is given as
\begin{equation}
    k_i = \tau\frac{ \sin\big( \frac{\pi(y-1)}{2}\big)}{2c_0}\omega^y +O((\tau\omega^{y-1})^3),
\end{equation}
and so in terms of the absorption parameters, setting $k_i=\alpha_0\omega^y$
\begin{equation}\label{tau}
    \tau=\frac{2 c_0 \alpha_0}{\sin\big( \frac{\pi(y-1)}{2}\big)}   .
\end{equation}
(For an extended derivation see \Rev{}{the} Supplementary Material.)

\subsection{Static-memory fractional time derivative}
This section summarises the theories of   \citet{birk2010improved} and \citet{gutleb2023static}, describing the method used to approximate the fractional time derivative.\Rev{We approach the construction similarly to that explained by \citet{diethelm2008investigation}, but making use of the updated quadrature rules given by \citet{birk2010improved}.}{} We begin by considering the Caputo fractional derivative of order $Y$ \cite{caputo1966linear} as given by 
\begin{subequations}
\begin{align}
    \pfrac{^\ty f}{t^\ty}(t)&=\frac{1}{\Gamma(\ceil{\ty}-\ty)}\int_0^t \frac{ f^{\ceil{\ty}}(\Tilde{t})}{(t-\Tilde{t})^{\ty+1-\ceil{\ty}}} \intd \Tilde{t},\\
    &=\frac{1}{\Gamma(1-\ty)}\int_0^t \frac{ f^{\prime}(\Tilde{t})}{(t-\Tilde{t})^{\ty}} \intd \Tilde{t},
\end{align}
\end{subequations}
where the latter equation arises because in this work we will only be concerned with $y=\ty+1\in(1,2)$, \textit{i.e.} $\ty\in (0,1)$, and so $\ceil{\ty}=1$. $f^\prime(t)$ denotes the ordinary derivative with respect to $t$.
In order to evaluate this integral we can expand the Gamma function into an integral between $(0,\infty)$ and apply a change of basis. This is detailed in the \Rev{Supplementary}{Supplementary} material.
\begin{subequations}\label{DoubIntFracDer}
    \begin{align}
    \pfrac{^\ty f}{t^\ty}(t)=&  \int_0^\infty \frac{2\sin(\pi \alpha)}{\pi} s^{2\ty-1}\phi_f(s,t) \intd s, \\
    \phi_f(s,t) =&  \int_0^t{\ex^{-s^2(t-\Tilde{t})} f^{\prime}(\Tilde{t})} \intd \Tilde{t} 
\end{align}
\end{subequations}
We proceed by evaluating the outer integral between $(0,\infty)$ through a quadrature rule, writing the problem as
\begin{equation}\label{sumtimefrac}
     \pfrac{^\ty f}{t^\ty}(t)=\sum_{j=1}^L A_j \int_0^t {\ex^{-s_j^2(t-\Tilde{t})} f^{\prime}(\Tilde{t})} \intd \Tilde{t}.
\end{equation}
To do this we will make use of the Gauss-Jacobi quadrature points $x_j$ and weights $\lambda_j$ for the interval $(-1,1)$ which results in \eqref{DoubIntFracDer} being expressed as,
    \begin{subequations}
    \begin{align}
        \pfrac{^\ty f}{t^\ty}(t)=&\sum_{j=0}^L \lambda_j\frac{8\sin(\pi \alpha)}{\pi (1+x_j)^4}\phi_f\lr{\frac{(1+x_j)^2}{(1-x_j)^2},t} \\
        \pfrac{^\ty f}{t^\ty}(t)=&\sum_{j=0}^L A_j\phi_f\lr{s_j,t}
        \end{align}
        with, 
            \begin{align}
                A_j= \lambda_j\frac{8\sin(\pi \alpha)}{\pi (1+x_j)^4}, \quad
                s_j= \frac{(1+x_j)^2}{(1-x_j)^2}.\label{AjSj}
            \end{align}
    \end{subequations}
We refer to $s_j$ and $A_j$ as the Birk-Song quadrature points and weights respectively, as described in \citep{birk2010improved}. $L$ is a free parameter dictating the number of quadrature points used\Rev{, with increasing $L$ increasing accuracy}{}. This process can be repeated for $Y>1$, adjusting \eqref{DoubIntFracDer} as given in \cite{birk2010improved, gutleb2023static}. Alternatives to the Birk-Song points are also available, such as those given in \cite{diethelm2008investigation} and \cite{yuan2002numerical}.

The advantage of writing the time-fractional derivative as a sum of integrals \Rev{over the original integral}{}is that $\phi_f$ lends itself easily to being updated in discrete time steps. Set
\begin{align}
    \phi_f(s_j,t)=&\psi_j(t)=\int_0^t{\ex^{-s_j^2(t-\Tilde{t})} f^{\prime}(\Tilde{t})} \intd \Tilde{t} \\
    \psi_j(t+\delta t) =& \ex^{-s_j^2\delta t} \psi_j(t) + \int_t ^{t+\delta t} \ex^{-s_j^2(t+\delta t-\Tilde{t})}f^\prime(\Tilde{t})\intd \Tilde{t}.
\end{align}
For suitably small $\delta t$ it is then possible to approximate $f^\prime(\Tilde{t})$ through interpolation, allowing the integral to be expressed in a closed form, which is exact for this interpolation. \Rev{For example, if we assume that}{Assuming}$f(t)$ is piece-wise linear on intervals $(n\delta t, (n+1)\delta t)$, it follows that
\begin{align}\label{psiupdate}
    \psi_j((n+1)\delta t) &= \ex^{-s_j^2\delta t} \psi_j(n \delta t) \nonumber \\ & \phantom{=}\,\,+\frac{1-\ex^{-s_j^2\delta t}}{s_j^2}\frac{f((n+1)\delta t)-f(n\delta t)}{\delta t}.
\end{align}
This has an error of order $O(N \delta t)$ after $N$ time steps in each of the $\psi_j$. Taking higher order interpolating polynomials also yields exact forms through integration by parts. \Rev{ However, we do not detail these further here as for the time step sizes we consider we get sufficient accuracy, and increasing the order of interpolation also increases the required memory of the method.}{This is detailed in the Supplementary material, though not used here due to the time step sizes considered.}

\subsection{Numerical scheme}
Here we describe a numerical procedure for evaluating \eqref{firstorder} in one dimension, using $\tau$ given by \eqref{tau} and evaluating the time-fractional derivative by \eqref{sumtimefrac}.\footnote{In order to make the comparisons between the numerical methods applied for the computation of the time-fractional loss given here and for the fractional-Laplacian method, which is implemented in the k-Wave toolbox \cite{treeby2010k}. The code used to evaluate the problem numerically is identical to that in k-Wave except for the treatment of the loss.} With the FFT denoted $\F$ and inverted by $\iF$. Additionally;
\begin{subequations}
    \begin{align}
        u^n =& u((n-1\slash 2)\delta t, \boldsymbol{x}),\\
        \rho^{n}=&\rho(n\delta t, \boldsymbol{x}),\\
        p^{n}=&p(n\delta t, \boldsymbol{x}),\\
        \psi^{n}_j=&\psi_j(n\delta t, \boldsymbol{x}).
    \end{align}
\end{subequations}
Where each variable is represented as an array of the size of the spatial grid corresponding to $\boldsymbol{x}$. (Note that the $\psi_j^n$ are $L$ such variables of spatial grid size, as indexed by $j$.) The update equations for acoustic particle velocity and pressure can be written as
\begin{subequations}\label{urhonumerical}
\begin{align}
    u^n=&u^{n-1}  -\frac{\delta t}{\rho_0}\iF\bigg( \mi k \kappa \F \big( p^n\big) \bigg)\\
    \rho^n=&\rho^{n-1}-\delta t\iF\bigg( \mi k \kappa \F \big( \rho_0 u^n\big) \bigg)\label{rhonumerical},
\end{align}
\end{subequations}
where $\kappa$ is a k-space correction to reduce numerical dispersion \cite{treeby2010k}.
Noting that $\tau$, $A_j$ and $s_j$ can be pre-computed using \eqref{tau} and \eqref{AjSj}, we then update the values of $\psi_j(n\delta t,\boldsymbol{x})$ of $\rho$ following equation \eqref{psiupdate}.
    \begin{align}\label{psinumerical}
    \psi_j^n&= \ex^{-s_j^2\delta t} \psi_j^{n-1} +\frac{1-\ex^{-s_j^2\delta t}}{s_j^2}\frac{\rho^n-\rho^{n-1}}{\delta t}.
\end{align}
Instead of the forward difference approximation,
we could have used the gradient as already computed using the FFT for \eqref{rhonumerical}. While for this specific case the result would be more accurate, removing any cancellation \Rev{areas}{errors}, when a perfectly matched layer (PML) is used this results in a reduced accuracy if not accounted for in the $\psi_j$ correctly. Since in these cases $\rho(n\delta t)$ has already been computed, and the cost for storing $\rho((n-1)\delta t)$ is fixed, we choose to use the form given by  \eqref{psinumerical}. Additionally, using the forward difference approximation allows for the inclusion of additive sources, which are added after the FFTs have been applied.
Now we are ready to close the system by computing $p^n$;
\begin{equation}\label{pnumerical}
    p^n=c_0^2 \bigg( \rho^n + \tau \sum_{j=1}^L A_j \psi_j \bigg)
\end{equation}
Where $\cdot_j$ represents the vector dot product in $j$ at each point in the grid. 
As previously mentioned, in the numerical results provided below we have additionally included a PML to prevent wave wrapping within the domain which is assumed to be periodic due to the use of the FFT. This change only impacts \eqref{urhonumerical} and is detailed within the literature, such as in \cite{tabei2002k,treeby2010modeling,treeby2010k}.

\section{Numerical Experiments. \label{sec:3}}
Here the numerical scheme consisting of \eqref{urhonumerical}, \eqref{psinumerical} and \eqref{pnumerical} will be demonstrated with examples \footnote{All of the numerical results presented in this paper have been produced using an adjusted version of the k-wave toolbox in MATLAB and is available at \cite{King2024TFGIT}.}. First, we will examine the accuracy of the time-fractional  numerical scheme as $L$ and $\delta t$ are varied. For the variation of $\delta t$ we make use of the \Rev{CFL number}{Courant–Friedrichs–Lewy (CFL) \cite{courant1967partial} number for 1D}, acting as a normalised time step given by CFL$= c_0 \delta t \slash \delta x$. As such, increasing or decreasing the CFL is equivalent to increasing or decreasing $\delta t$ respectively. (With $c_0$ and $\delta x$ remaining fixed.)\Rev{}{ Traditionally, the CFL number is used to ensure stability, with an explicit time solver as we have here, being stable for CFL$<1$. In higher dimensions the formulation sums the above expression across each of the dimensions, however for simplicity the number we give here will always be given only by the first dimension. \\}
Second, the absorption power-laws that result from the scheme are compared to those that use the fraction Laplacian, as \Rev{}{implemented }in k-Wave. 
Finally, 1D and 3D examples are given in which the absorption power-law exponent, $y$, varies throughout the domain.
Throughout this section we make use of the parameters given in Table \ref{tab:param}, unless stated otherwise in the figure captions.
\begin{table}[h!]
\caption{Typical parameters for $c_0$, $\alpha=\alpha_0 \omega^y$ and $\rho_0$. Figures \ref{fig:MaxErr}-\ref{fig:VariedAbsorption1D} used (a)-(c), and the named tissues were used for Figure \ref{fig:13}. The data are taken from [1] \cite{treeby2010modeling}, [2] \cite{lou2017generation}, [3] \cite{szabo2004diagnostic} Table B.1., [4] \cite{moran1995ultrasonic}}.
\centering
\begin{ruledtabular}\label{tab:param}
\begin{tabular}{l|ccccc}
& $c_0$ & $\alpha_0$ & $y$ & $\rho_0$ & Cite \\
\hline
(a)& 1500 & 0.5 & 1.1 & 1 & [1]\\
(b)& 1500 &0.25 & 1.5 & 1 & [1]\\
(c)& 1500 & 0.1 & 1.9 & 1 & [1]\\
Fat& 1470 &  0.6 & 1.01 & 937 & [2,3]\\
Blood vessel& 1584 & 0.14& 1.21 & 1040 & [2,3]\\
Skin & 1650 & 0.22 & 1.15 & 1150 & [2,4]\\
Fibroglandular & 1515 & 0.75 & 1.5 & 1040 & [2,3]\\
\end{tabular}
\end{ruledtabular}
\end{table}

\subsection{Convergence in $L$ and $\delta t$.}

\begin{figure}[h!]%
    \centering
    \includegraphics[width=.5\textwidth]{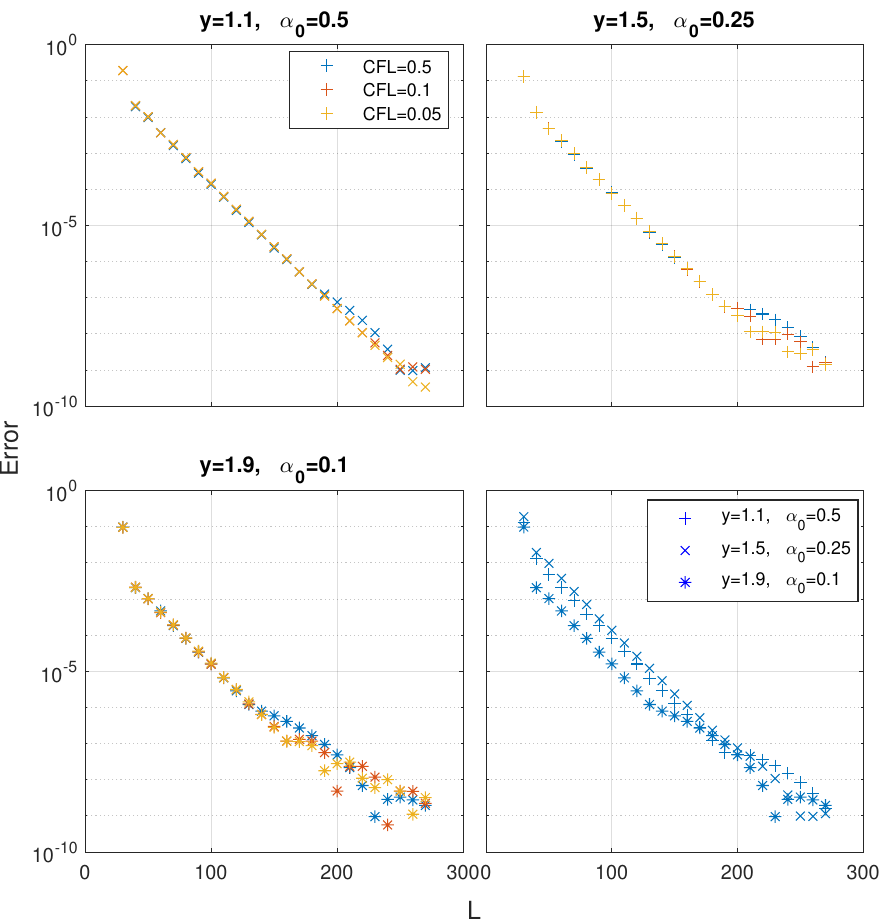}
    \caption{Normalised maximum error, $l^\infty$, of the pressure at the sensor between $p(t,L)$ and $p(t,L-10)$. Plotted for parameter sets (a), (b) and (c) as indicated according to Table \ref{tab:param} with CFL=0.5, 0.1 and 0.05, and for CFL=0.5 for each parameter set (bottom right).}
    \label{fig:MaxErr}
\end{figure}

\begin{figure}[!ht]%
    \centering
    \includegraphics[width=.5\textwidth]{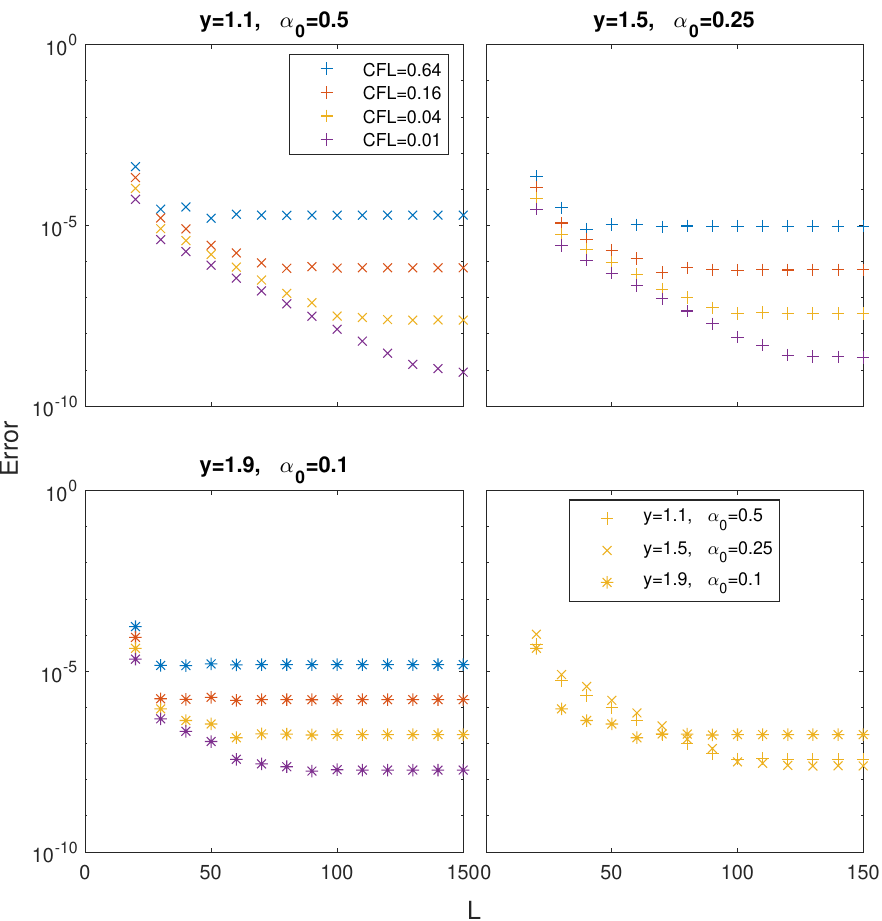}
    \caption{As shown in FIG. \ref{fig:MaxErr} comparing p(t,L) with the indicated CFLs to p(t,300) with a CFL of $6.25\times 10^{-4}$. Plotted with CFL=0.64, 0.16, 0.04, and 0.01 and for CFL=0.04 for each $y$ (bottom right).}
    \label{fig:MaxrefErr}
\end{figure}

For Figures \ref{fig:MaxErr}\Rev{,\ref{fig:NormErr}}{} and \ref{fig:MaxrefErr}, three different power-laws $\alpha=\alpha_0 \omega^y$ and three different timesteps $\delta t$ (through the CFL) are considered. The problem, a 1D simulation across $16mm$ with a point source located $4mm$ into the domain, is run over 7 microseconds with a sensor located $10mm$ from a point pressure source. The errors considered are those in the sensor signal.
In Figure \ref{fig:MaxErr}\Rev{ and \ref{fig:NormErr}}{} this is done sequentially, comparing the signals for each fixed value to $L$ to the signal for $L-10$, with Figure \ref{fig:MaxErr} plotting the error in the $l^\infty$ norm \Rev{, and Figure \ref{fig:NormErr} making use of the $l^2$ norm. It is observed that the choice of error does not change the overall behaviour, only the scale, all other errors are computed using just the $l^\infty$ norm.}{with the $l^2$ norm behaving similarly.}

\Rev{In contrast to Figure \ref{fig:MaxErr} and \ref{fig:NormErr},}{} Figure \ref{fig:MaxrefErr} instead compares the signals to a reference produced by running the same simulation with a CFL of $6.25\times 10^{-4}$ and \Rev{an L value }{}$L=300$. In order to compare the reference signal to the sensor signals the reference signal is sampled at the times for which the sensor records a point. \Rev{The error is plotted using the $l^\infty$ norm. The CFLs chosen in Figure \ref{fig:MaxrefErr} were chosen specifically in order to quarter the time step size between each case.}{}

In Figure \ref{fig:MaxErr}\Rev{and \ref{fig:NormErr}}{} it can be observed that irrespective of the time step size the improvement in the sensor output is uniform between the case for increasing values of $L$. This remains true even for different values of $y$ with each case also giving similar results.
Despite the consistent convergence in $L$ for any given CFL, this does not mean we are converging to an exact solution, rather for large enough $L$ the error from the choice of $\delta t$ becomes dominant. As a result, for each $\delta t$ we converge to a solution with a minimum fixed error in $L$. This same behaviour was observed in \citep{gutleb2023static} for the approximation of the fractional derivative. This is most clearly seen here in Figure \ref{fig:MaxrefErr} where for each of the CFL considered, increasing $L$ decreases the error only so far before it plateaus and becomes approximately constant. For a decreased CFL, and therefore smaller time step $\delta t$, it may be observed that a higher value of $L$ is required to reach this plateau, however for each fixed $L$ decreasing the time step results in a smaller error.

\begin{figure}[!ht]%
    \centering
    \includegraphics[width=.5\textwidth]{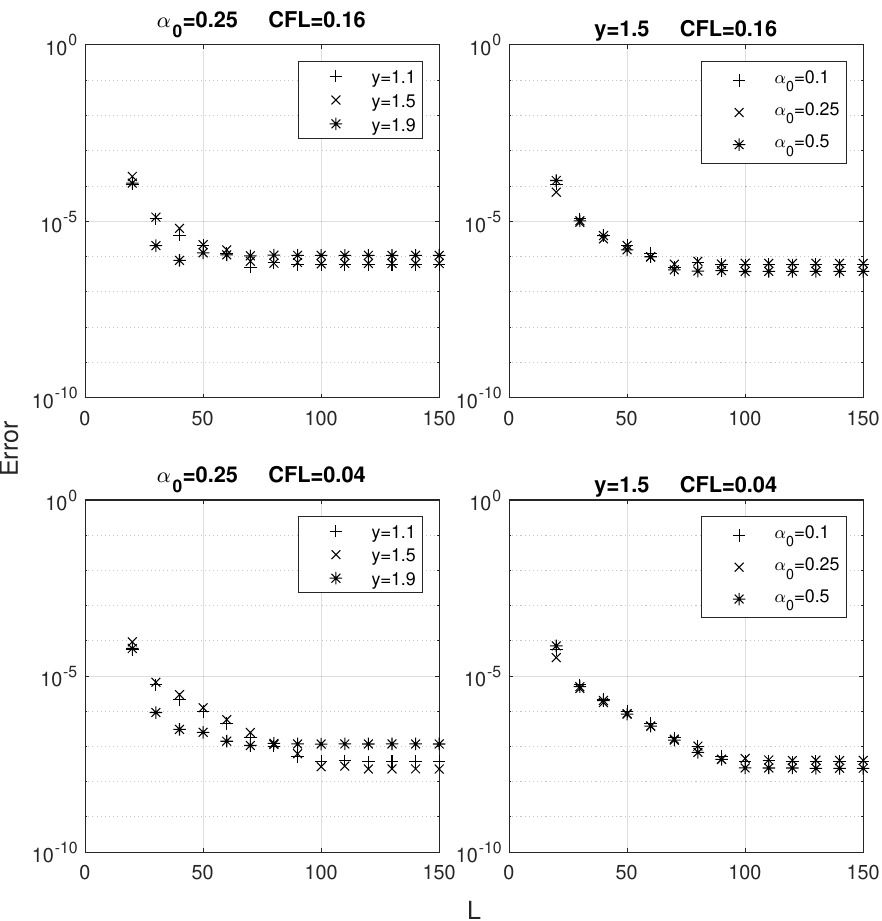}
    \caption{As FIG. \ref{fig:MaxrefErr} comparing p(t,L) with the indicated CFLs to p(t,300) with a CFL of $6.25\times 10^{-4}$. \Rev{Plotted with CFL=0.64, 0.16, 0.04, and 0.01 and for CFL=0.04 for each $y$ (bottom right) $\alpha_0=0.25$ is the same for all cases.}{Plotted with CFL=0.16 (top) and CFL=0.04 (bottom) we  vary $y$ for a fixed $\alpha_0=0.25$ (left), and vary $\alpha_0$ for a fixed $y=1.5$ (right).}}
    \label{fig:MaxErry}
\end{figure}

While it can be observed that each of the three $y$ values produce the similar error irrespective of the choice of the CFL for large $y$, In Figure \ref{fig:MaxrefErr} we may observe that the rate at which the error plateaus is different, becoming steady for a smaller value of $L$ for $y$ closer to $2$, suggesting that for the larger values of $y$ the error for a fixed $\delta t$ is larger compared to the error from the choice of $L$, and therefore dominates. This is more clearly \Rev{seen between Figures \ref{fig:MaxErry} and \ref{fig:MaxErra0}}{ seen in Figure \ref{fig:MaxErry}} where a fixed $\alpha_0$ and varied $y$, and a fixed $y$ and varied $\alpha_0$ are considered. \Rev{ the same errors as for Figure \ref{fig:MaxrefErr} are plotted.}{}

In Figure \ref{fig:MaxErry} \Rev{}{(left)} it can be observed that, when just $y$ is varied,\Rev{, although all the results are very similar for a large CFL value, the behaviour changes for smaller values of the CFL, with the larger values of $y$ having a larger error. This is particularly visible in Figure \ref{fig:MaxErry}(d).}{ most of the behaviour of Figure \ref{fig:MaxrefErr} (bottom right) is still captured. With the larger values of $y$ producing smaller errors for the same value of $L$ until the error plateaus.}
In Figure \ref{fig:MaxErry} \Rev{}{(right)} on the other hand the three \Rev{plots for the}{} different values of $\alpha_0$ present very little variation, showing the exact same behaviour across the different CFL and $L$ values. While the larger values for $\alpha_0$ do perform marginally better across the cases, this difference is not comparable to the differences observed in Figure \ref{fig:MaxrefErr}\Rev{ and \ref{fig:MaxErry} which exhibit very similar behaviours, with both varying the value of $y$}{}.

Rather than comparing the time-fractional solution to itself we can also compare against the Green's function solution to the second order PDE problem \eqref{SecondOrder}. \Rev{}{Approximate methods for computing the Green's function are given in for example; \citet{kelly2016approximate}.} Alternatively, we can make use of the exact Green's function solution given in \citet{treeby2011k}, which instead of the time-fractional derivative term gives the loss in terms of the fractional-Laplacian. It should be noted that the two problems are equivalent up to the order of perturbation within the linearised Euler equations with loss. The advantage of using the Green's function solutions are that they are exact when considered in homogeneous domains. \Rev{While these solutions cannot easily be adapted to model heterogeneous problems, may provide a suitable benchmark. As a base case we are additionally able to compare}{While the Green's function solution cannot be extended to the heterogeneous model, when considered in a homogeneous domain it provides a suitable test case, capturing the correct absorption behaviour. Additionally we may compare} the rates of convergence between the \Rev{k-Wave code, which makes use of the fractional-Laplacian}{fractional-Laplacian loss}, and the time-fractional approach \Rev{. Both approaches, with their different loss functions, tend to the exact solution given by the second order Green's function,}{to the Green's function,} which is included in the k-Wave MATLAB package (Mathworks Inc., Natick, MA)\cite{treeby2011k}.

\begin{figure}[h!]%
    \centering
    \includegraphics[width=.5\textwidth]{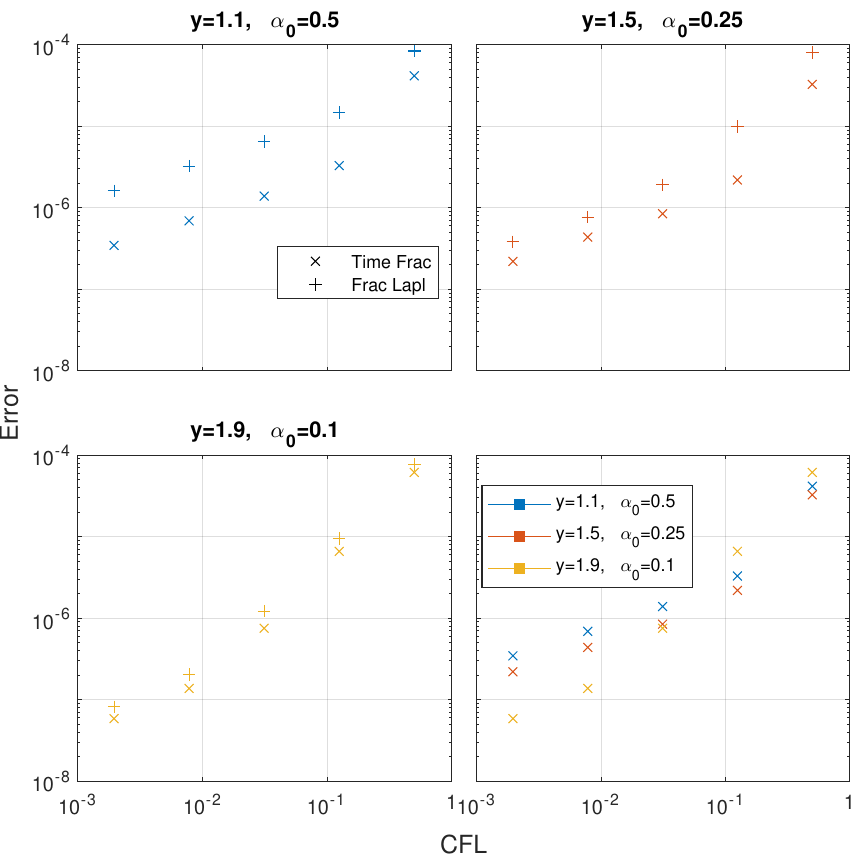}
    \caption{ Comparison between the reference Green’s function solution for the
fractional-Laplacian loss and the solution from the first order equations
with the time-fractional loss term ($L=80$, $\times$) and with the fractional-
Laplacian loss ($+$). Errors computed for CFL numbers between $2^{-1}$ and $2^{-9}$. The errors have been normalised against the maximum value of thesecond order solution at the sensor location. Pairs of $\alpha_0$ and $y$ are as used in figure \ref{fig:MaxErr}, with the bottom right plot only showing the results for the time-fractional  derivative for each power-law considered.}
    \label{fig:ErrSecond}
\end{figure}%

Plotted in Figure \ref{fig:ErrSecond} are the normalised errors for the first order ordinary differential equation (ODE) solver, with loss described both by the fractional-Laplacian and the time-fractional derivative as discussed here, with L=80. These are compared to the second order ODE Green's function for the fractional-Laplacian loss\Rev{, which is exact in a homogeneous domain. $L$ was chosen as 80}{. The choice of L is due to the
near-convergence seen for each of the CFL numbers} in Figures \ref{fig:MaxrefErr} and \ref{fig:MaxErry}, with the error dictated by choice of the CFL and not the quadrature rule, except for exceptionally small time steps. Improved errors may be observed for the smaller CFL numbers by increasing $L$ further. \Rev{}{This choice of $L$ additionally limits the increased computational cost of the method, both in memory storage, and number of operations compared to the fractional-Laplacian method, with $L$ increasing these both linearly. This is detailed further in section \ref{sec:5}.}

For all three of the parameter pairs for $\alpha$ it may be observed that both methods converge to the Green's function solution \Rev{to the second order problem with fractional-Laplacian loss, }{}as would be expected, confirming that in the small $\delta t$ limit the two methods are equivalent, further validating the time-fractional method as an alternative approach to modelling power-law absorption to the traditional fractional-Laplacian method. In addition to \Rev{both methods converging,}{this} it is interesting to notice that despite representing a different loss term, the model that uses the time-fractional derivative converges faster than the fractional-Laplacian model to the \Rev{true solution}{Green's function solution with fractional-Laplacian loss}. In particular, between $CFL=2^{-3}$ and $2^{-9}$; in \Rev{case (a)}{the case of $y=1.1$ and $\alpha_0=0.5$} the time-fractional  method produces an error of the same magnitude as the fractional-Laplacian with a CFL $2^{-4}$ smaller, i.e. the same accuracy is being retrieved for a time step that is 16 times larger. For the \Rev{$\alpha$ value used in (b)}{values $y=1.5$ and $\alpha_0=0.25$} this is reduced to only 4 times larger, and \Rev{case (c)}{in the final case} only observes a small improvement. However, noting that for soft tissues $y$ is typically close to 1, implies that the improvement seen in \Rev{case (a)}{the first case} could represent a significant computational saving by considering the time-fractional  approach. \Rev{It should additionally be observed through (d) that the loss of improvement between the methods for larger $y$ values coincides with an overall increase in accuracy.}{Observing the bottom right plot of Figure \ref{fig:ErrSecond}, which plots all three cases just for the fractional-Laplacian method, as the CFL decreases the error decreases at a faster rate in $y$. This coincides with the reduced difference between the time-fractional and fractional-Laplacian methods which already allows for a larger time step.}

\Rev{Finally, it should be noted that 
the error converged to the limit set by $\delta t$ faster for larger values of $y$. This suggests that there is an increased likelihood that these error bounds could not be improved as much by further increasing $L$ for the smaller values over case (c).}{}

\subsection{Absorption}
\begin{figure}[h!]%
    \centering
    \includegraphics[width=.5\textwidth]{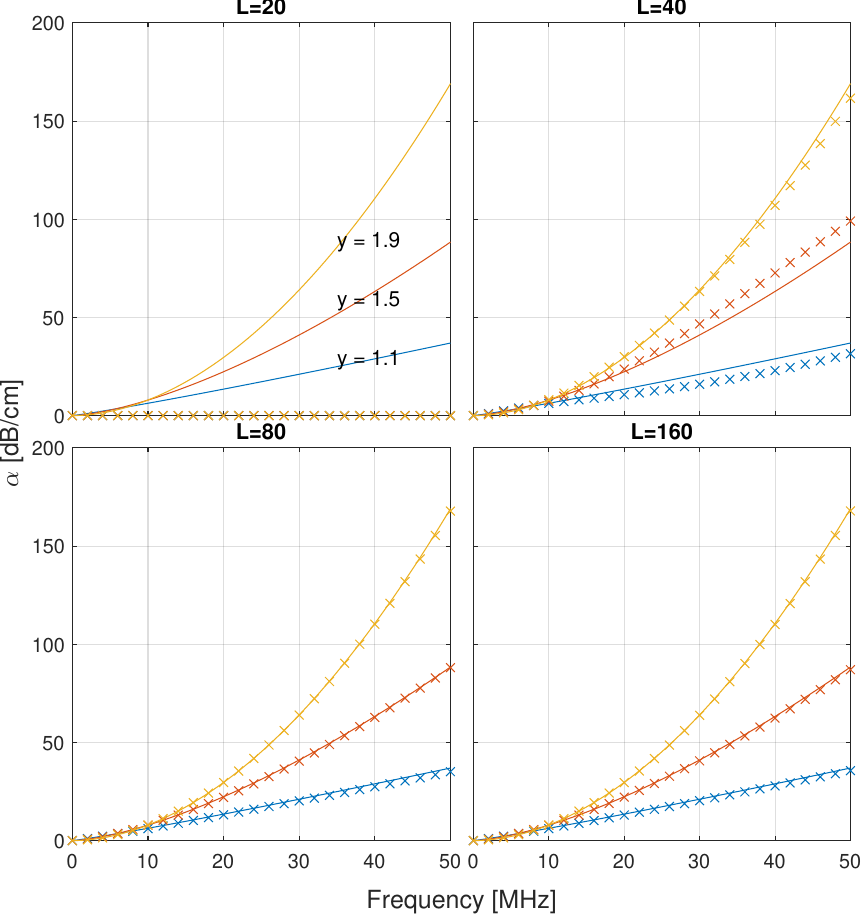}
    \caption{Absorption alpha for \Rev{L=30,60,120,180}{$L=20$, $40$, $80$, $160$} computed as described in the main text. Solid lines describe the predicted absorption while circles mark on the values found through the simulation. For $L=20$ (top left), the
results from the predicted absorption are indistinguishable between the
three cases.}
    \label{fig:alphaL}
\end{figure}

\begin{figure}[h!]%
    \centering
    \includegraphics[width=.5\textwidth]{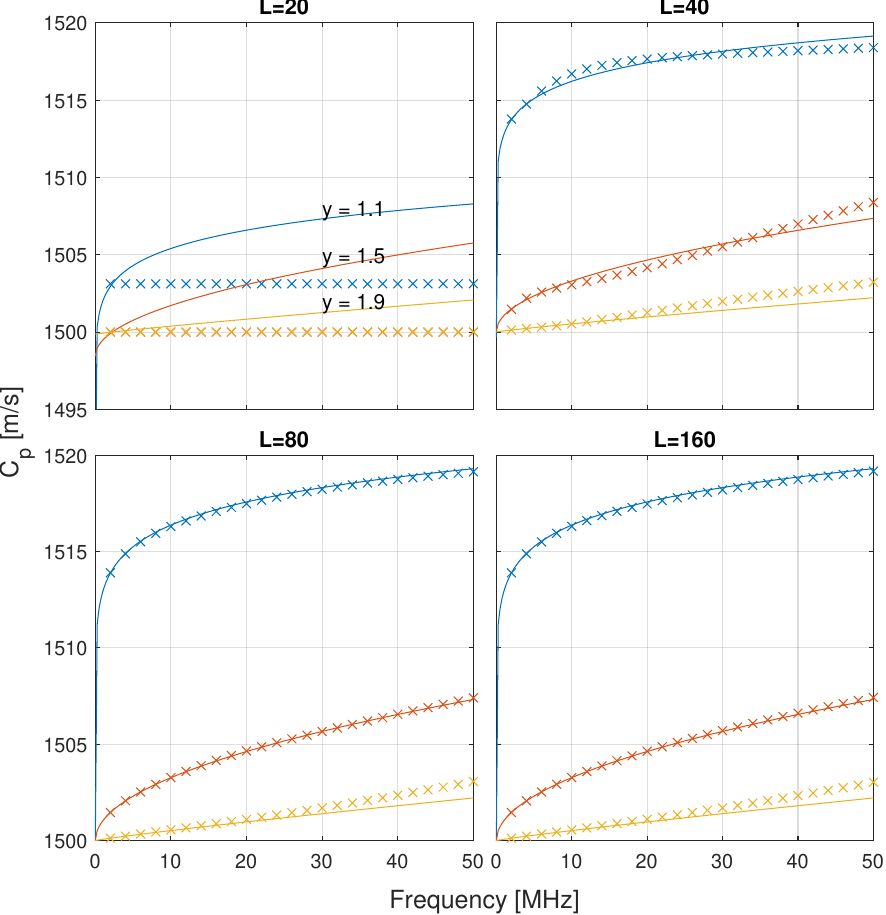}
    \caption{As Figure \ref{fig:alphaL} but showing the dispersion. Both predicted (solid lines) and from the simulation.}
    \label{fig:cpL}
\end{figure}

\Rev{Considering the same power-law absorption as in the previous section w}{W}e now test how well the method reproduces each $\alpha(\omega)$ by comparing the observed absorption to the frequency-dependant power-law given by the input parameters. In addition to this we make use of the Kramers-Kronig relation \cite{waters2005causality,waters2000applicability} to compare the observed dispersive wave speed to predicted values for the given power-law. \Rev{This observes both how closely $\alpha$ is re-produced in the frequency domain and the phase speed $c_p$.}{} This is done as in \cite{treeby2010modeling} for the loss generated by the fractional-Laplacian \Rev{, making use of the k-Wave toolbox, }{}which we will additionally compare against.
Each of the Figures in this section have been produced considering the following simulation: A 1D 12.2mm domain is considered under a uniform spatial grid of size 1024, over a time of 4 microseconds for a CFL of 0.05. A point pressure source applied at time t=0 is located 3.05mm into the domain, with two sensors located 3.55mm and 4.55mm in the domain. The sensor data is recorded \Rev{at each of these sensors }{}and their respective amplitude and phase spectra are found. This information is then processed into the absorption by comparing the amplitude spectrum of the two sensors, and the dispersion through the phase spectrum. Figures $\ref{fig:alphaL}$ and $\ref{fig:cpL}$ plot $\alpha$ and $C_p$ for four values of $L$. These are $L=20$, $L=40$, $L=80$, and $L=160$.
For $L=20$ it is clear that the simulation has failed to predict the absorption correctly, with no frequency dependence observed, but rather a constant value of $\alpha$. This problem is mostly resolved for $L=40$, however each case observes divergence between the prediction and simulations as the frequency increases. Increasing to $L=80$ reduces the error for all three with only small differences between the cases of $L=80$ and $L=160$.
This is similarly observed for the dispersion, with $L=20$ unable to correct predict $c_p$, while $L=40$ reproduces the prediction for low frequencies but then breaks down for the higher frequencies. Similar results are  produced by $L=80$ and $L=160$; however for $y=1.9$ the higher frequencies still observe this difference. This may be anticipated through Figure \ref{fig:MaxrefErr} where in the bottom right plot we compare the errors for the three values of $y$. Here we observe that, while for $y=1.9$ we see convergence in L faster than for the lower values of $y$, it is observable that these lower values actually produce a smaller error for a CFL of 0.04, which is comparable to that used for this simulation.

\begin{figure}[h!]%
    \centering
    \includegraphics[width=.5\textwidth]{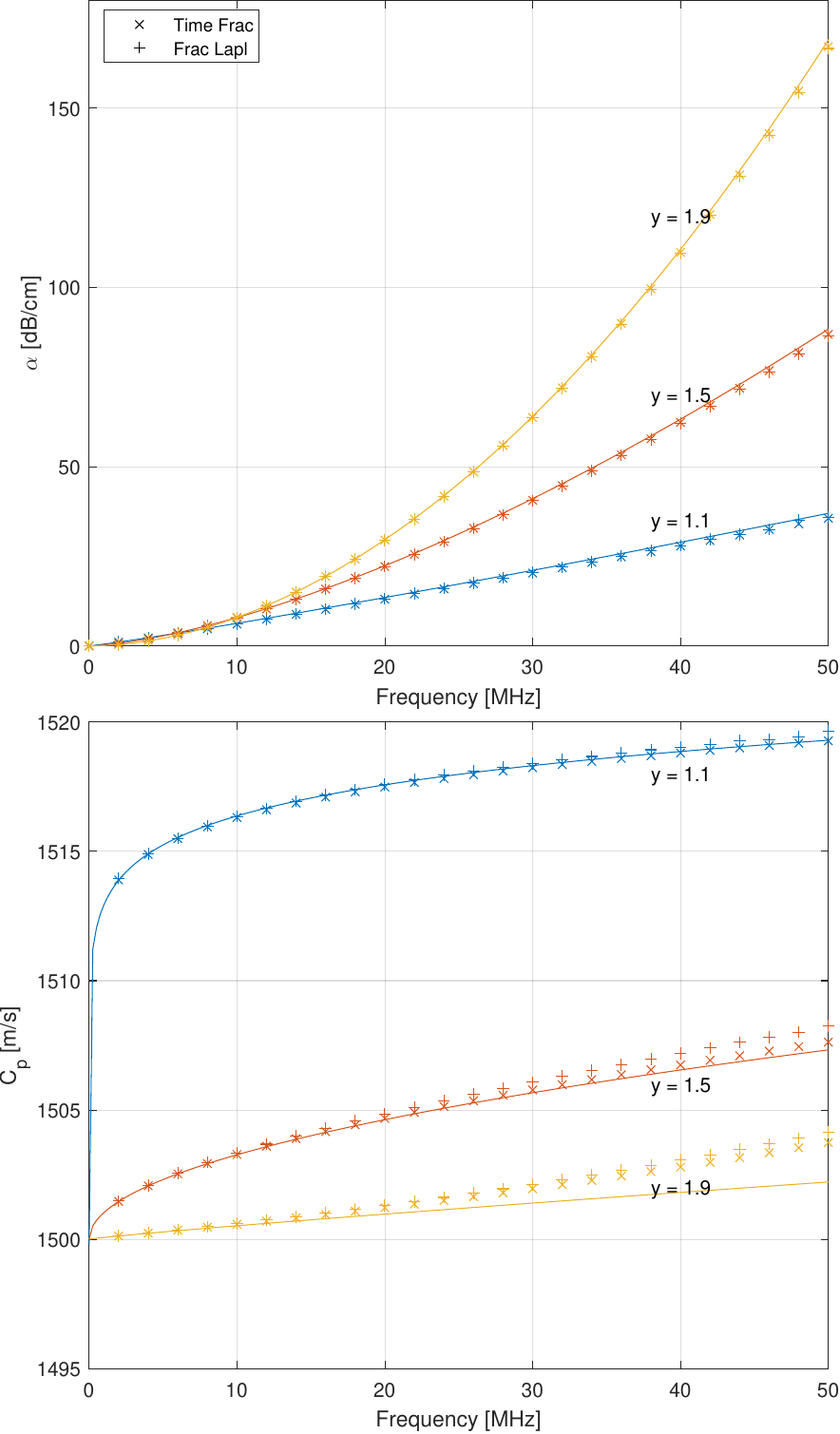}
    \caption{As Figures \ref{fig:alphaL} (top) and \ref{fig:cpL} (bottom) plotting the alpha for both the time-fractional  derivative ($\times$) with $L=120$, and with the fractional-Laplacian ($+$).} 
    \label{fig:FracComp}
\end{figure}

It is natural to then compare these results to those produced for the fractional-Laplacian using the k-Wave toolbox, this is performed in Figure \ref{fig:FracComp}. Here we have chosen $L=120$.
In this figure it can be observed for each of the values of $y$ that the results produced are very similar between the two methods. However it may be observed in the dispersion for $y=1.5$ and $y=1.9$, where the time-fractional derivative was least accurate to the predictions, that this method performs better than when the loss is given by the fractional-Laplacian. This improvement from the fractional-Laplacian to the timefractional
derivative may have been anticipated from Figure \ref{fig:ErrSecond} where this improvement was observed for each parameter set
considered. \Rev{}{Observing the plot of the dispersive sound speed it may be suggested that the larger values of $y$ present a larger error, suggesting a contradiction to the results of Figure \ref{fig:ErrSecond}. There are two important features to note: First the waves with the highest frequency will decay fastest, which is where this error is observed most readily, minimising the effect of these high frequency regions at the sensor locations. And second, even when these errors are larger, the time-fractional method performs better than the fractional-Laplacian method.}

\subsection{Spatially-varying power-law absorption}\label{sect:varalpha}

One advantage of the time-fractional  approach is that, unlike the fractional-Laplacian, the derivative is local in space. Heterogeneities typically manifest in three forms within biological media, the ambient density $\rho_0$, the sound speed $c_0$ and the absorption $\alpha$. The variation in $\rho_0$, which manifests as $\boldsymbol{u}\cdot\boldsymbol{\nabla} \rho_0$ within the conservation of \Rev{momentum}{mass} equation, can be removed, reducing the equations back to the homogeneous form, by considering $\rho$ not as the true density but as the adjusted parameter described by $\rho-\boldsymbol{d}\cdot\boldsymbol{\nabla} \rho_0$ where $\boldsymbol{d}$ is the particle displacement. As such the acoustic density fluctuation is not computed directly. There is in addition an adjusted loss term, however the difference is small in the majority of cases, and zero where $\boldsymbol{\nabla} \rho_0=0$, and therefore may be neglected in the majority of cases. A varied $c_0$ is only observed in the computation of the equation of state, appearing both in the pre-factor, and within the loss term.\\
It is not common numerically to account for a varied absorption \Rev{. This is despite it's obvious presence within the application}{despite being known to influence the final result within applications}. For the time-fractional method, since each location in the domain is treated locally in the equation of state, we need only pre-compute the $A_j$ and $s_j$ \eqref{AjSj} parameters for each location. Proceeding exactly as before, with increased storage, but no increased computational cost for each time step.

\begin{figure}[h!]%
    \centering
    \includegraphics[width=.5\textwidth]{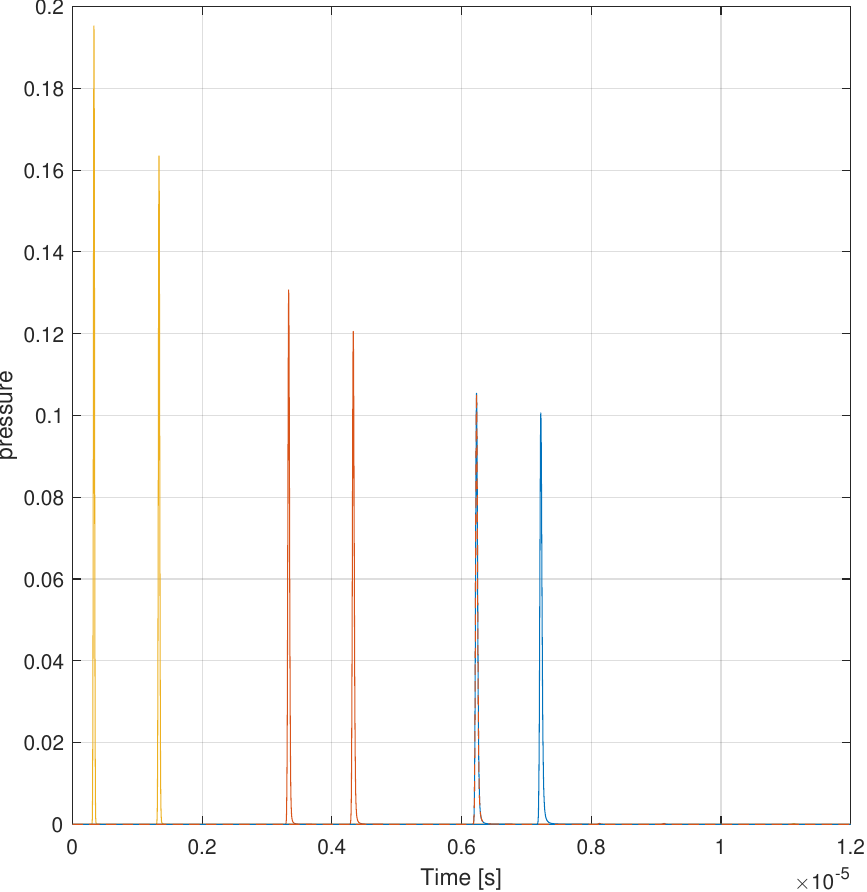}
    \caption{Plot of the pressure at each time for each of the six sensors, with colours indicating the region of absorption in which they are located. \Rev{Black}{Yellow} plots correspond to $\Rev{\alpha_0=0.5}{\alpha_0=0.25}$ and \Rev{$y=1.1$}{$y=1.9$} as between sensors one and two. Red plots correspond to \Rev{$\alpha_0=0.1$}{$\alpha_0=0.125$} and \Rev{$y=1.9$}{$y=1.5$}, as used between sensors three and five. Blue plots use \Rev{$\alpha_0=0.5$, $y=1.5$}{$\alpha_0=0.25$, $y=1.1$}, as for sensors five and six. Note that sensor five is located on the \Rev{boarder}{border} between two regions.} 
    \label{fig:VariedSensor1D}
\end{figure}%

\begin{figure}[h!]%
    \centering
    \includegraphics[width=.5\textwidth]{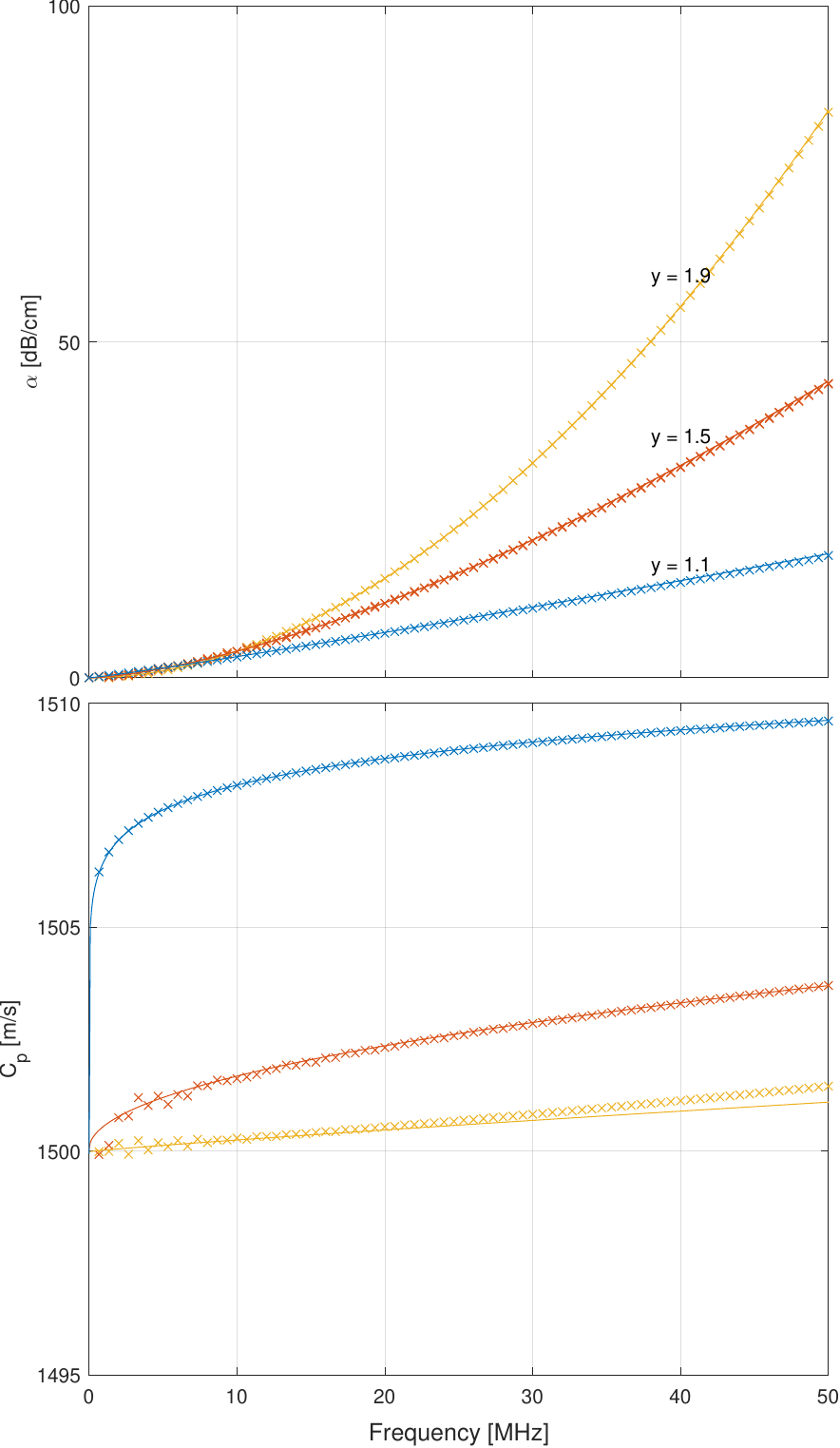}
    \caption{As Figure \ref{fig:FracComp} plotting $\alpha$ and dispersion from the time-fractional  derivative with $L=100$, with absorption plotted between sensors; one and two, three and four, four and five, and, five and six. \Rev{Black}{Yellow} plots correspond to \Rev{$\alpha_0=0.5$ and $y=1.1$}{$\alpha_0=0.05$ and $y=1.9$} as between sensors one and two. Red plots correspond to \Rev{$\alpha_0=0.1$ and $y=1.9$}{$\alpha_0=0.125$ and $y=1.5$}, as used between sensors three and five. Blue plots use \Rev{$\alpha_0=0.5$, $y=1.5$}{$\alpha_0=0.25$, $y=1.1$}, as for sensors five and six.} 
    \label{fig:VariedAbsorption1D}
\end{figure}%

In Figures \ref{fig:VariedSensor1D} and \ref{fig:VariedAbsorption1D} a system with varied absorption is illustrated. We have considered a 1D, 18.8mm domain with 2048 grid-points. A point source is located at 4.7mm \Rev{into the domain}{from the left edge of the domain}. Sensors have been placed 5.2, 6.2, 8.2, 9.2, 11.1 and 12.1mm \Rev{from the source}{ to the right of the point source}. We have then enforced that the absorption is defined by $\alpha=0.05\omega^{1.9}$ between 0mm and 7.2mm \Rev{}{from the left edge of} the domain, and $\alpha=0.125\omega^{1.5}$ between 7.2mm and 11.1mm. \Rev{in the domain, and}{And} finally $\alpha=0.25\omega^{1.1}$ from 11.1mm to the \Rev{end}{rightmost edge} of the domain. This locates \Rev{}{the first two sensors within the first region, the third and forth sensors in the second region, and finally the sixth sensor in the third region.} The fifth sensor \Rev{}{is located} exactly at the interchange between the second and third region of the domain.
\Rev{}{It should be noted that the $\alpha_0$ values used here are half of the values given in Table \ref{tab:param}. This reduces the absorption effects and reflections as the wave changes between the regions and as a result, the numerical errors introduced in the calculation of $\alpha$ and $c_p$ from the numerical data.\\}
In Figure \ref{fig:VariedSensor1D} the sensor data from each of the six sensors can be observed, with colours indicating which region of the domain they are located in. These are then translated into plots of the absorption and the dispersion in Figure \ref{fig:VariedAbsorption1D} where matching colours have been used to indicate the regions for comparisons against the actual absorption and the Kramers-Kronig relation indicated once again with solid lines.

In Figure \ref{fig:VariedAbsorption1D} there is a region in which an error can be observed unalike those observed in Figures \ref{fig:alphaL} and \ref{fig:FracComp}. For $\omega <10 MHz$ for $y=1.5$, observed more clearly in the plot of the dispersive sound speed, the Kramers-Kronig relation is poorly predicted. This is a result of a reflected waves being produced due to the change in absorption, since these propagate in the opposite direction to the initial wave, despite their small amplitude. This same error occurs in the first region, $y=1.9$ however it is less noticeable.

\begin{figure}[h!]%
    \centering
    \includegraphics[width=.49\textwidth]{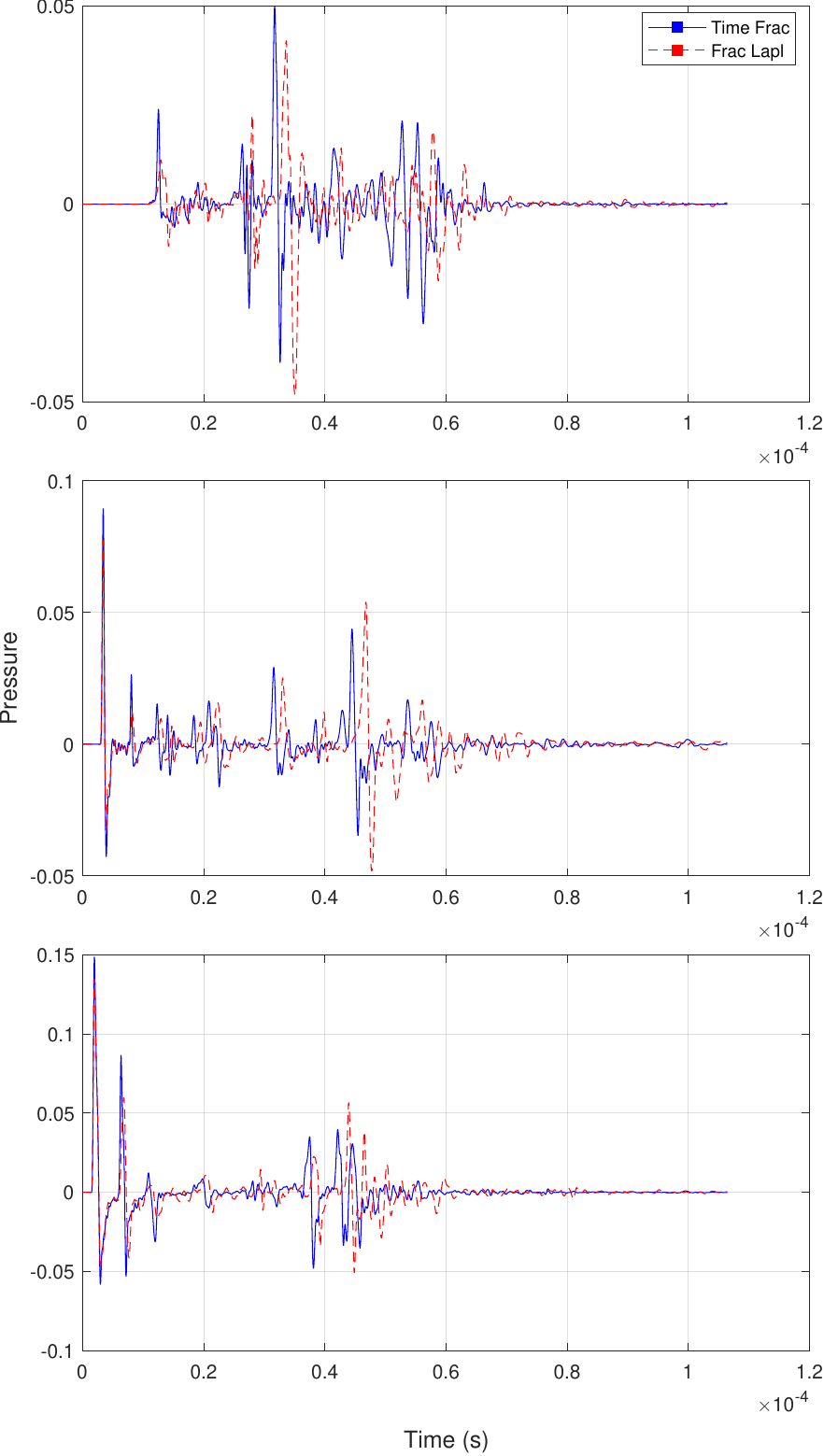}
    \caption{Plots of three sensor data of the pressure at locations outside the breast. Plotted: (Red) homogeneous absorption with the fractional-Laplacian method. (Blue) Heterogeneous absorption with the time-fractional  method.}
    \label{fig:13}
\end{figure}%

It is additionally possible to model a varied absorption in both 2D and 3D, with Figure \ref{fig:13} this is illustrated for a 3D example. We make use of a 3D Breast Phantom provided alongside \citet{lou2017generation}. In the breast phantom we have modelled the different tissue types with different sound speeds and densities. In addition we have then applied a varied absorption according to the tissue type when running the simulation with the time-fractional  method, or a uniform absorption for a loss described by the fractional-Laplacian. Sensors were located on the ``outside" of the breast such that only the effects of the breast material properties are observed with a source represented by a single pulse from each voxel representing blood vessels at time $t=0$.

From Figure \ref{fig:13} it can be observed that while both simulations produce similar signals, as time progresses, the arrival of the signals gradually vary an increased amount, while from even the first arrival time the amplitude of the waves are different. Which signal is greater, however, is not consistent with time. This variation between the signal  with greater amplitude highlights that the two signals are decaying at different rates. While it should be noted that the homogeneous case would illustrate some additional errors, since the same CFL has been used for both simulations and the homogeneous cases uses the k-Wave toolbox with the fractional-Laplacian loss which was seen to have greater error in Figure \ref{fig:ErrSecond}. It is clear that the differences between the two signals is mostly produced by the difference in the absorption, while the absorption used in the homogeneous case is chosen to match the absorption observed in most of the domain.  As such the waves will travel through regions with both higher and lower absorption with earlier arriving waves mainly travelling through the skin, and the later travelling through more regions of fat which has a lower absorption power-law exponent.

\section{\label{sec:5}\Rev{Conclusion}{Discussion}}
\Rev{Presented in this paper is an alternative approach to modelling power-law absorption can be modelled in the wave equation. This is performed by making use of the Caputo time-fractional derivative. \Rev{The co-efficient for this derivative has been derived and details of the approximations used are given both analytically and numerically. Additionally, this}{ This} method has then been graphically illustrated to correctly predict the absorption for suitably chosen free parameters $L,\delta t$. Finally comparisons have been made numerically to power-law absorption modelled by the fractional-Laplacian derivative. This is through the running of the k-Wave software package, which makes use of the fractional-Laplacian loss, as well as an adjusted version with changes only to the loss operator to encode the time-fractional loss given here.}{}

The time-fractional static memory approach \Rev{used here has}{to the Caputo time-fractional derivative given in this paper has} been previously detailed by \citet{gutleb2023static} where it was compared against a variety of model problems in order to illustrate the effectiveness of the method. In their work the fractional time derivative was applied to a second order differential equation for acoustic loss, although this is not the same acoustic loss as used in this work, even when transformed into the second order equations. Further distinguishing the works is the approach taken to the spatial dimensions, with the work here applying the time-fractional derivative on a uniform spatial grid. This may be considered a loss in problem generality, however it allows the method to be more directly compared to the aforementioned numerical implementations  of the fractional-Laplacian while also allowing easy treatment of the \Rev{spacial}{spatial} domain.

The Caputo fractional derivative is defined by an integral over the full problem history. In order to avoid having to store the problem history within the numerical implementation, as is used for the static memory and logarithmic memory principles, here the problem history is stored in an iteratively defined $\phi_\rho(t)$, which when approximated numerically only requires a single time step of the problem history.\Rev{This approximation has been labelled $\psi$ and is generated through a linear interpolation of $\rho(t)$ between time $t$ and $t+\delta t$.}{} Storing the problem history in this way does introduce a second integral on $(0,\infty)$ for which a quadrature rule has been applied with variables relating to the $j^{th}$ quadrature point indicated with a subscript $j$, with the total number of quadrature points denoted $L$, which is free to be chosen within the implementation of the method.

For a fixed value of $L$ and computational grid of size $N_X$, the resulting method for approximating the fractional derivative requires a fixed memory cost, which is why this method is referred to as a static memory method. Applying the fractional time derivative to compute the loss operator only requires the additional storage of $L(N_X + 2) + N_X$  scalar variables compared to the memory requirements of the k-Wave toolbox having changed only the method for computing the loss.\Rev{ This additionally accounts for recording one step of the history of $\rho$. }{}This is a large memory requirement, particularly for 3D computational grids, given that L would typically be chosen between 80 and 120 in order to have likely observed convergence in $L$ such that the remaining error comes from the time step, see Figures \ref{fig:MaxrefErr}, \ref{fig:MaxErry}. It is noted however that for many problems the total number of time steps to be taken would be much larger than L, and storing the whole problem memory would require $n \times (N_X + 1)$ variables after $n$ time steps.

The alternative ‘short term memory’ and ‘logarithmic memory' principles choose which points of the problem history to be stored reducing this cost, however both require a long history, increasing memory requirements or lose accuracy as the total number of time steps increases. Of these methods the logarithmic memory principle is more accurate for the same computational cost, while the fixed memory principle is as good or more accurate for a fixed length of history being considered.

In light of this it is worth considering current HPC trends where it may be observed that systems continue to evolve to address memory-bound applications. As such methods like the one presented here may become increasingly attractive for large-scale acoustic simulations. The time-fractional approach presented here offers potential performance benefits for certain computing architectures, despite its increased memory requirements compared to the fractional-Laplacian method. While the additional memory usage may be problematic for compute-bound systems like GPUs, it is well-suited for emerging CPU designs optimized for memory-intensive workloads. For example, the NVIDIA Grace CPU Superchip (Nvidia, Santa Clara, CA) provides up to 960 GB of high-bandwidth LPDDR5X memory, making it capable of handling the increased memory footprint of this method. Additionally, by eliminating the need for the additional FFT operations per time step, this approach reduces computational complexity, which could lead to improved performance on memory-bound systems, particularly for the varied absorption case.

In works such as \cite{chen2004fractional} it is suggested that the fractional-Laplacian lends itself more directly to modelling variation in the spatial dimensions, being physically more valid. However, since the operator is derived globally by spatial Fourier transforms, local changes to the operator are not easy to account for, as such the\Rev{ official implementation of k-Wave}{ k-Wave toolbox} only allows for a single power-law absorption. 
\Rev{In contrast, u}{U}sing a time-fractional loss allows spatial changes in the derivative to not impact the computation, since each location has the corresponding fractional time derivative computed locally. This results in the additional computational cost of the method being reduced to the pre-computation of the $s_j$ and $A_j$ values for each value of y, with the trade off being that the memory requirement is increased to $(3L + 1)N_X$. In cases where the \Rev{spacial}{spatial} grid is large and only a few values of $y$ are being considered , it would be possible to reduce the storage to $L(N_X + 2Y ) + N_X$ where $Y$ is now the number of different power-laws. However, this would have the increased computational cost of $Y$ sums over $j$ of $A_j \psi_j$ . These methods were used in the generation of Figures \ref{fig:VariedAbsorption1D} and \ref{fig:13} respectively. Considerations between which approach to use for a varied absorption power would include both storage and computation considerations, particularly if smoothing was performed on $y$ across the domain. In order to generate Figure \ref{fig:13} the computational time for the homogeneous case was approximately 5 times lower than that for the heterogeneous case.

In addition to the memory requirement and the observed computational time when comparing between loss computed with the fractional-Laplacian and the time-fractional methods for the loss also of importance is the accuracy. In Figure \ref{fig:ErrSecond} it is observable that the time-fractional method produced significantly lower orders of magnitude of error than that of the fractional-Laplacian loss for the same size of time step, particularly for values of $y$ close to 1. It is possible that as a result of this for such values a time step that is two to four times larger could be utilised producing the same error for the absorption. It has not been investigated if this difference in time stepping could suitably account for the computational differences.

\section{\Rev{}{Conclusion}}

\Rev{}{The method for computation of the Caputo time-fractional derivative as described \citet{birk2010improved} has been coded and applied to modelling power-law absorption within the first order wave equations as equivalent to the second order equation given by Caputo/Wismer \eqref{SecondOrder}. It has been verified that the method performs as good or better at predicting the power-law absorption and dispersive sound speeds as the methods describing the loss by the fractional-Laplacian method for the same size of time step under a suitably chosen $L$ which describes the number of Gauss-Jacobi quadrature points used.}

\Rev{}{The method has then been applied to allow for varied power-law absorption which has not previously been detailed within the literature. This is despite varied absorption being physically relevant to applications including for ultrasound absorption as the wave propagates through different tissue.}

Further work on this topic would include further optimisations to the formulations for $A_j$, $s_j$ and $\phi_j(t)$, including look up tables for the first two. It is however of note that the accurate calculation of the $A_j$ and $s_j$ is paramount to the method working, with higher point precision being required in the pre-computations performed in the time-fractional approach in order to retrieve an accurate approximation of the time-fractional derivative. 

Additionally, it is possible that alternative choices of the quadrature rule for approximating the first order derivative, $\pfrac{\rho}{t}$ within the computation of $\phi_j$ may allow for even larger time-steps, and exact forms of $\phi_j$ are given within the Supplementary material. However, preliminary tests of these methods have not yielded the expected improvements but performed worse, even on model problems. This is possibly due to cancellation errors for the small time steps such as the ones used on a fine grid, this would additionally raise the storage requirement of the problem.

It is possible that this method could be applied to other problems with fractional derivatives outside of the range $1<y<2$, as already detailed in \cite{birk2010improved} with only small changes being made to the specific loss term in use.

\Rev{}{Finally, comparisons could be made directly to existing methods for evaluation of the time-fractional derivative on the same model equation with spacial variations, such as the short term memory principle \cite{podlubny1998fractional}, the logarithmic memory principle, \cite{ford2001numerical} and an approximate method for the equivalent Green's function solution \cite{kelly2016approximate}.} 

\section{Supplementary Material}
See supplementary material at [URL will be inserted by AIP] for breakdowns of the mathematical formulations for $\tau$, the rewriting of the fractional derivative and integral evaluation under interpolation.

\begin{acknowledgments}
This research was supported by the Engineering and Physical Sciences Research Council (EPSRC) under grant Nos EP/W029324/1, EP/T022280/1, and EP/T014369/1. We offer thanks to Dr. Jiří Jaroš for his insights on the computational trends.
\end{acknowledgments}
\section*{Aurthor Declarations}
The authors have no conflicts to disclose.

\section*{Data Availability}
Data sharing is not applicable to this article as no new data were created or
analyzed in this study.

 
\bibliography{sampbib}
\end{document}